# A Randomized Generic Lucas Seed Algorithm (RGLSA) with Tail Boosting for Threat Modeling in Virtual Machines

**Abstract.**

This research paper will analyze security threats and proposes the self-propagating model of seeding attacks in cloud computing. Threat modeling on distributed and self-organizing systems is a very important modeling paradigm which discusses and analyzes the different ways malware may propagate in such systems. The paper introduces Randomized Generic Lucas Seed Algorithm (RGLSA) with Tail- Boosting which is based on Lucas and Fibonacci sequences. This is a model where the Virtual machines (hosts) could get infected rapidly by collaborative and recursive growth of the seeds generated in a random order. Tail boosting is introduced, for the first time so that in the scenario of the Cloud environment getting scaled up, the attack probabilities on VM's don't decrease drastically. The randomized growth of the seeds ensure that simulated attacks are free from a deterministic pattern and therefore, all the more challenging to be detected.

# 1. Introduction

## How important is Security and trust, anyway?

In a Cloud Environment, the customer divides his data among several service providers (*SP*'s) available in the market, based on his available budget. SP's provide a decision for the customer, to which *SP*s he must chose to access data, with respect to data access quality of service offered by the *SP*s at the location of data retrieval. This not only rules out the possibility of a *SP* misusing the customers' data, breaching the privacy of data, but can easily ensure the data availability with a better quality of service.

Infrastructure as a Service (IaaS) serves as the foundation layer for the other delivery models, and a lack of security in this layer will certainly affect the other delivery models, i.e., PaaS, and SaaS that are built upon IaaS layer [1]. From our point of view, the number one service or feature that is missing is security of data. There are two levels of concern here. One is focused on preventing others (such as another customer) from reading private data. This is a clear and obvious concern and prominent in scenarios such as theft, or other direct malicious attack. The other is concerned with the service provider reading private data. Besides simple lack of trust of the provider themselves, it should be obvious that the service provider is not 100% immune to attacks or other malicious activity, targeted or otherwise. These two levels of concerns apply to other security issues as well, and of course are commensurate with the level of confidentiality desired [2]. They considered intruder model and requirements that need to be satisfied to provide required level of privacy. Since previous research show that crypto- graphic means cannot always provide protection (especially in long term) they proposed a trust-based privacy protection. Their approach was based on subjective logic that applied to measure/monitor level of trustworthiness of cloud service providers. They explained how users have to handle their data to minimize privacy threats in the cloud [3].

As more and more information on individuals and companies are placed in the cloud, concerns are beginning to grow about just how safe an environment it is. It is better to prevent security threats before they enter into the systems and there is no way how this can be prevented without knowing where they originate. This brings about the necessity of security threat modeling [4, 5, 6, 7]. In this paper, we propose a scheme for threat modeling "Randomized Generic Lucas Seed Model" which may affect the virtual machines (VMs) on the cloud. The phrase "Cloud" originates from the cloud symbol used by flow charts and diagrams to symbolize the Internet. The term Cloud Computing refers to both the applications delivered as services over the Internet and the servers and system software in the datacenters that provide those services. The virtual machines (VMs) on the cloud will be affected due to the sharing of resources among themselves. Only virtual machine can affect the rest and that is why we need to analyze the attacks invading the VM so that we could prevent their spread in the entire cloud.

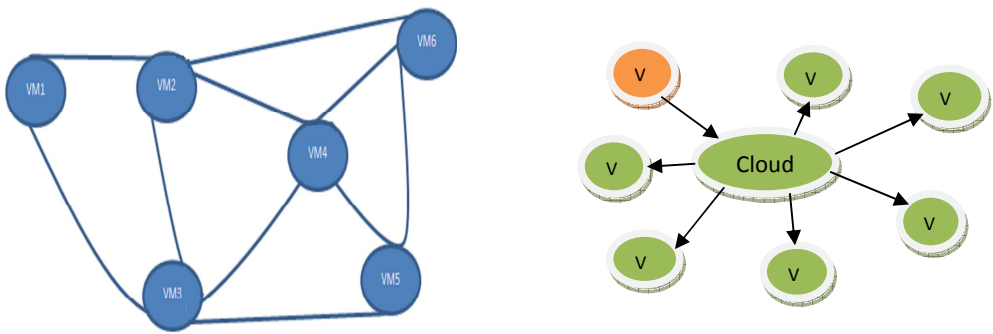

**Figure 1. A typical Cloud Set-up with the VM's pooling resources**

There are multiple stages in Malware detection and propagation. Scanning ports to discover vulnerable hosts are one of such, upon which malicious payload can be transmitted. Self- propagating seeds are quick to replicate, without human intervention and therefore have the ability to pose more serious threats to the network, Cloud or otherwise. Vulnerability in VM's need to be discovered and upon doing that, the final stage is to destroy/ modify resources of hosts/VM's. The process should be concurrent and be able to run on multiple ports/threads. There has been some work in this domain while most of the work concentrated on non-concurrent spanning of the worms/seeds.

The rest of the paper is organized as follows. Section 2 overviews the relevant work done in the domain briefly. Section 3 discusses our model with analytical validation. Section 4 contains experimental outcomes and discussions. Section 5 concludes our work.

## 2. Related Work

A discrete-time, deterministic Fibonacci Malware propagation model has been discussed in [8]. Analytical and computational estimations are based on discrete-fibonacci sequence with self-replicating seeding mechanism. In viral marketing, a key problem is to select an initial"seed" set from the network such that the entire network adopts any behavior given to the seed. Here a method was introduced for quickly finding sets of seeds that scale to very large networks. The tipping model approach found a set of nodes that guarantees flooding the entire network [9]. Various scanning algorithms have been proposed in the past, namely localized scanning, naïve random scanning, hit-list scanning, permutation scanning [10, 11, 12] Collaborative models of Malware propagation have been explored by researchers [13]. Detection architectures are endowed with [14] a virus throttle program that could detect abnormal network behavior. A Threshold random walk algorithm to identify remote, infected or potentially malicious hosts has been proposed by [15].

However, to the best of our knowledge, a randomized algorithm which will alter the growth pattern each and every time the simulation is run has not been done by anybody or any research group. Our algorithm is inspired by the behavior of Fibonacci and Lucas Sequences and randomizing the outcomes for growth and and attack probability optimization.

## 3. Spreading of Malicious Attacks on Cloud: Our Work

The proposed Malware attack model can pose a serious threat in a self-organizing network like cloud system. It has the ability to infect multiple hosts at a time. This can damage the host in several ways like consuming resources or giving false impression of high payload etc. Here we assume that $VM_1$ is the source of attack and initiation of malicious attack. Due to the sharing of resources, it affects any machine on the cloud call it physical machine. Once the physical machine is affected, there is a possibility that $VM_2$, $VM_3$....$VM_{n-1}$ might be affected and the process continues until the entire cloud is affected.

**Model Assumptions:**

1. The VM's are connected in such a way that so that the sequence $\{VM_1, VM_2, VM_3....VM_{n-1}\}$ can be followed initially. Parallel edges might exist for multiple seeding of the same instances.
2. If multiple seeds are generated at a certain time instant, then one seed attacks the nearest available VM in the sequence, other copies of seeds get added to the next seed value and attacks the next VM. The VM's are modeled as objects/obstacles which when encountered in the scanning path, raises the flag to 1, otherwise set to 0.
3. Define a 1-1 correspondence between the set of VM's, $\{VM_1, VM_2, VM_3....VM_n\}$ and the set of malicious seeds $\{L_0,...L_{n-1}\}$ initially. This guarantees that no seed is lost and multiple copies could be used for flooding almost instantaneously.

### 3.1. Proposed Randomized Lucas seed model system

Assume that at least one VM is affected; virus is seeded onto other VMs. The dynamic sharing of resources among VMs allows each VM to be affected by viruses and eventually affects the entire cloud. In this model, we use Random Fibonacci sequences. This can be demonstrated below:

Given that:
$$L_n = \frac{1}{\gamma}(a_{n-1} + a_{n+1}) \quad \text{(i)}$$
$$L_0 = 2\gamma, \quad L_1 = \gamma$$

are the first pair of seeds. $L_n$ is the possible number of nodes to be affected. Now, $a_n$ is a solution

(well-known) to the deterministic difference equation below:

$$a_{n+1} = a_{n-1} + a_n, \; n \geq 2$$
$$a_0 = 0, \; a_1 = 1$$

from eqation (1) above, we assume that All VMs talk to each and $\gamma$ is the Random Fibonacci sequence.

Properties of Random Fibonacci sequences :

1. $\sum_{i=1}^{n} F_i^2 = F_n * F_{n+1}$

2. $F_n = \frac{1}{\sqrt{5}} \frac{(1+\sqrt{5})^n}{2} + \frac{(1-\sqrt{5})^n}{2}$

3. $L_n = L_{n-1} + L_{n-2}, \; n \geq 2$     (ii)
   $L_0 = 2\gamma, \; L_1 = \gamma$

$\frac{L_j}{L_n} \equiv$ the probability that VM "j" may be affected given that all previous VMs are affected. It is the transmission probability that node j will be affected. Here node and VM are used interchangeably.

### 3.2 Guiding Assumptions for seed propagation:

1. Initially (at $0^{th}$ instance) no VMs are affected.

2. $L_n \equiv$ Possible number of VMs to be affected.

3. $\frac{L_i}{L_n}$ =Transmission probability.

4. Assume "$\gamma$" is the random number between 0 & ½ that a VM request the same resource be attacked. $L_n$ is a solution to the difference equation

$$L_{n+1} = \frac{1}{\gamma}(L_n + L_{n-1}), \; n \geq 2$$
$$L_0 = 2, \; L_1 = 1 \quad \text{and}$$
$$L_n = \frac{1}{\gamma}(a_{n-1} + a_{n+1}) \quad \text{where}$$

$$a_n = \frac{\gamma}{\sqrt{5}} \{ \frac{(1+\sqrt{5})^n}{2} - \frac{(1-\sqrt{5})^n}{2} \}$$

is a solution to

$$a_{n+1} = \frac{1}{\gamma}(a_{n-1} + a_{n+1})$$
$$a_0 = 0, \; a_1 = 1$$
$$\gamma \in random\, numbers(0,1)$$

### 3.3 Analytical Justification for self-replication:

Lemma 1: $\frac{\gamma}{\sqrt{5}} \left( \frac{1+\sqrt{5}}{2} \right)^n - \frac{\gamma}{\sqrt{5}} \left( \frac{1-\sqrt{5}}{2} \right)$ solves the equation $a_{n+1} = \frac{1}{\gamma}(a_{n-1} + a_n)$

$where, a_0 = 0, a_1 = \gamma$

**Proof:** $n = 0, n = 1$ true (simple substitution by inspection)

*hypothesis:* if $n = k$ is true, then $a_{k+1} = a_{k-1} + a_k$

*Induction step:*   if NTS is true for $n = k+1$,

then $a_{n+2} = a_k + a_{k+1}$

$$= \frac{1}{\sqrt{5}}\left\{\left(\frac{1+\sqrt{5}}{2}\right)^k - \left(\frac{1-\sqrt{5}}{2}\right)^k\right\} + \frac{1}{\sqrt{5}}\left\{\left(\frac{1+\sqrt{5}}{2}\right)^{k+1} - \left(\frac{1-\sqrt{5}}{2}\right)^{k+1}\right\}$$

$$= \frac{1}{\sqrt{5}}\left\{\left(\frac{1+\sqrt{5}}{2}\right)^k\left(1+\frac{1+\sqrt{5}}{2}\right) - \left(\frac{1-\sqrt{5}}{2}\right)^k\left(1+\frac{1-\sqrt{5}}{2}\right)\right\}$$

$$= \frac{1}{\sqrt{5}}\left\{\left(\frac{1+\sqrt{5}}{2}\right)^k\left(\frac{2+1+\sqrt{5}}{2}\right) - \left(\frac{1-\sqrt{5}}{2}\right)^k\left(\frac{2+1-\sqrt{5}}{2}\right)\right\}$$

$$= \frac{1}{\sqrt{5}}\left\{\left(\frac{1+\sqrt{5}}{2}\right)^k \frac{1}{2^2}(6+2\sqrt{5}) - \left(\frac{1-\sqrt{5}}{2}\right)^k \frac{1}{2^2}(6-2\sqrt{5})\right\}$$

$$= \frac{1}{\sqrt{5}}\left\{\left(\frac{1+\sqrt{5}}{2}\right)^k\left(\frac{1+\sqrt{5}}{2}\right)^2 - \left(\frac{1-\sqrt{5}}{2}\right)^k\left(\frac{1-\sqrt{5}}{2}\right)^2\right\}$$

$$= \frac{1}{\sqrt{5}}\left\{\left(\frac{1+\sqrt{5}}{2}\right)^{k+2} - \left(\frac{1-\sqrt{5}}{2}\right)^{k-2}\right\}$$

RHS $= a_{k+2}$

(Proved)

Next, using the randomization component

$$a_n = \frac{\gamma}{\sqrt{5}}\left\{\left(\frac{1+\sqrt{5}}{2}\right)^n - \left(\frac{1-\sqrt{5}}{2}\right)^n\right\}$$

$a_0 = 0, a_1 = \gamma$

RHS

$= a_k + a_{k+1}$

$$= \frac{\gamma}{\sqrt{5}}\left\{\left(\frac{1+\sqrt{5}}{2}\right)^k - \left(\frac{1-\sqrt{5}}{2}\right)^k\right\} + \frac{\gamma}{\sqrt{5}}\left\{\left(\frac{1+\sqrt{5}}{2}\right)^{k+1} - \left(\frac{1-\sqrt{5}}{2}\right)^{k+1}\right\}$$ and continuing in the same fashion, we obtain

$= \gamma a_{k+2}$

$$= \frac{\gamma}{\sqrt{5}}\left\{\left(\frac{1+\sqrt{5}}{2}\right)^k - \left(\frac{1-\sqrt{5}}{2}\right)^k\right\}$$

and, $a_0 = 0, a_1 = \gamma$

Thus, the initial seeding mechanism is validated by the above proven theorem, which justifies the solution to the following difference equation.

$$a_{n+1} = (a_{n-1} + a_n)\frac{1}{\gamma}$$

$$while, a_0 = 0 \,\&\, a_1 = \gamma$$

$$\gamma \in rand(0,1)$$

- **Theorem 2:** $\gamma\left\{\left(\frac{1+\sqrt{5}}{2}\right)^n + \left(\frac{1-\sqrt{5}}{2}\right)^n\right\}$ solves the recurrence relation $L_{n+1} = \frac{1}{\gamma}(L_n + L_{n-1}), n \geq 2$

$L_1 = \gamma, L_0 = 2\gamma$ where $\gamma \in rand(0,1/2)$ and $L_n = \frac{1}{\gamma}(a_{n-1} + a_{n+1})$ where $a_n$ is a solution to $\frac{L_i}{L_n} \equiv$ probability that VM 'i' may be affected given all previous VM's are affected $\vartheta_i S_L \to S_{VM}$ such that $\vartheta$ is i-1.

$$\{S_L \equiv \text{Set which contains the Lucas number } L_1, L_2\ldots\ldots L_n\}$$

$$\{S_{VM} \equiv \text{Set which contains the VM's } V_1, V_2\ldots\ldots V_n\}$$

Proof: ( By induction) $L_0, L_1$ checks easily. For n=k, assume the hypothesis to be true.

i.e $L_{k+1} = \frac{1}{\gamma}(L_k + L_{k+1}); k \geq 2$ is true. NTS, n=k+1 is true $L_{k+2} = \frac{1}{\gamma}(L_{k+1} + L_k)$

$$L_{k+1} + L_k = \gamma\left\{\left(\frac{1+\sqrt{5}}{2}\right)^{k+1} + \left(\frac{1-\sqrt{5}}{2}\right)^{k+1}\right\} + \gamma\left\{\left(\frac{1+\sqrt{5}}{2}\right)^k + \left(\frac{1-\sqrt{5}}{2}\right)^k\right\}$$

$$= \gamma\left\{\left(\frac{1+\sqrt{5}}{2}\right)^k\left(\frac{1+\sqrt{5}}{2}+1\right)\right\} + \gamma\left\{\left(\frac{1-\sqrt{5}}{2}\right)^k\left(\frac{1-\sqrt{5}}{2}+1\right)\right\}$$

$$= \gamma\left\{\left(\frac{1+\sqrt{5}}{2}\right)^k \frac{1}{2^2}(6+2\sqrt{5}) + \left(\frac{1-\sqrt{5}}{2}\right)^k \frac{1}{2^2}(6-2\sqrt{5})\right\}$$

$$= \gamma\left\{\left(\frac{1+\sqrt{5}}{2}\right)^k\left(\frac{1+\sqrt{5}}{2}\right)^2 + \left(\frac{1-\sqrt{5}}{2}\right)^k\left(\frac{1-\sqrt{5}}{2}\right)^2\right\}$$

$$= \gamma\left\{\left(\frac{1+\sqrt{5}}{2}\right)^{k+2} + \left(\frac{1-\sqrt{5}}{2}\right)^{k+2}\right\}$$

$$= \gamma L_{k+2}$$

Hence, $L_n = \frac{1}{\gamma}(a_{n-1} + a_{n+1})$ follows similarity.

## Conclusion of the Theorem:

1. The growth of the seeds, defined by the Lucas sequence are recursive in nature, hence self-replicating.

2. The growth of the Lucas seeds is exponential.

3. In case the Cloud is scaled up which is a very common scenario, the transmission probabilities may go down since the denominator increases while the numerator doesn't. Tail boosting by using an altered transmission probability heuristic = $\dfrac{L_i + L_j}{L_{n+j}}$ is initiated under the condition that the new probability <= 1.

    Note: "j" is an arbitrary number of VM's added to the system under attack.

    Interpretation: $L_0$, $L_1$ are the first pair of seeding attack and the subsequent attacks are modeled by the prescribed algorithmic approach –Randomized Generic Lucas Seed Algorithm (RGLSA) as discussed below.

### Randomized Generic Lucas Seed Algorithm (RGLSA)

| | |
|---|---|
| I. Accept the number of VM in the existing cloud network, identified as n. | 1. Assume there are n numbers of VM existing in a cloud environment. A malicious attack of virus is affecting one target VM in the entire cloud. |
| II. Create a for loop i→n, continues till n number of iteration.<br><br>III. Define Lucas function inside the for loop as lucas(i).<br><br>   a. Define a time function which calculates the time to generate Lucas value. | 2. To detect the probability factor how a VM affects. To obtain the transitional probability we have already proved that the following function has given rise to a new seeding mechanism, the generated seed is represented by a mapping function such as $a_{n+1} = \dfrac{1}{\gamma}(a_{n-1} + a_n)$, Where $a_0 = 0$ and $a_1 = 1$ and $\gamma$ is a random number generated as $\gamma \in rand(0,1)$ |
| IV. call lucas() function<br>   a.  generate a gamma factor between(0,1) and multiply with 0.5.<br>   b.  create an alpha factor which accept the reverse gamma.<br>   c.  For lucas value $L_0 \& L_1$ the transitional probability is defined as 1 for both.<br>   d.  For lucas vale $L_2......L_n$ define recursive Lucas relation as<br>      $L_n = alpha * (Fibo_{n-1} + Fibo_{n+1})$.<br>   e.  Define an array to store the values of Lucas factor $L_1, L_2.............L_n$<br>   f.  Now define transitional factor as $L_i / L_n$ where $(0<i<n)$.<br>   e. Define a new variable which store the summation factor(sum) of $L_1, L_2.............L_n$<br>V. Call Fibo()<br>  a. generate a gamma factor between(0,1).<br>  b. generate an alpha factor which takes reverse gamma.<br>  c. For Fibonacci value $Fibo_0, Fibo_1$ consider value 0 and 1 respectively.<br>  d. For Fibonacci value $Fibo_2,.............Fibo_n$ calculate recursive relation as<br>  alpha*$(Fibo_{n-1} + Fibo_{n+1})$ | 3. For any positive integer value of n number of VM, the system calls a Lucas function which represent as $L_n$, representing the $i^{th}$ number of VM to be affected( where $0<i<n$) $L_n = \frac{1}{\gamma}(a_{n-1} + a_{n+1})$ ,while $L_0 = 2$ and $L_1 = 1$ (which can be proved) And $L_0, L_1$ is the 1st pair of seeding attack.<br>4. Now every $a_i$ in Lucas function is defined by a Fibonacci series number which reclusively generated and summarized as the $L_n$ .<br>5. To obtain the ultimate transitional probability for every VM we calculate $\frac{L_i}{L_n}$ considering n as the upper bound of VM.<br>-----The transitional probabilities at the tail of Lucas sequence by adding a random number alpha (0,0.5) to the numerator of transitional probability formula $\frac{L_i}{L_n}$ So that always $\frac{L_i}{L_n} < 1$ This will help the malware propagation in the event that VMs are scaled( as the n increases). |

## 4. The Experiment:

The simulation, set up in JAVA and GNUPlot is expected to achieve the following goals.
i) The growth of malicious seeds on VMs ($L_n$ versus n)- numerical validation of the theory using randomization.

ii) $\frac{L_i}{L_n}$ Vs. n- Observe the transmission/attack probability growth in a dynamic and scalable environment and to observe how "Tail-boosting" helps adjust the sagging probabilities and ensure that the VM's are consistently attacked.

iii) $L_n$ Versus. Time (milliseconds)—to observe how quickly the seeds grow and transmit.

iv) "Tail-boosting" to control the infection and damage caused by the malicious seeds.

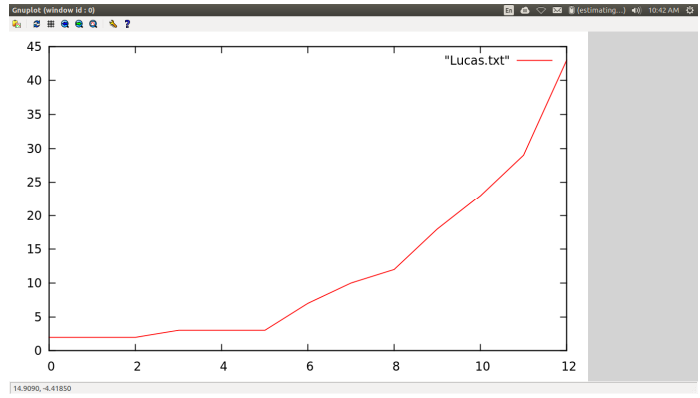

**Figure 2: The growth of malicious seeds on VMs ($L_n$ versus n) for one seed-set value**

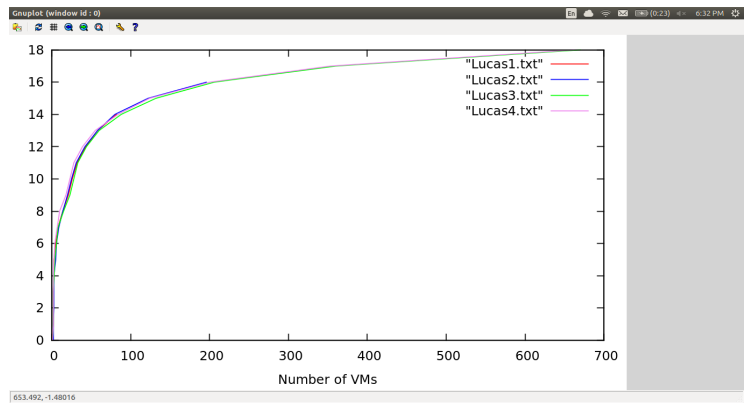

**Figure 3: The growth of malicious seeds on VMs ($L_n$ versus n) for n=4, 8, 10, 12**

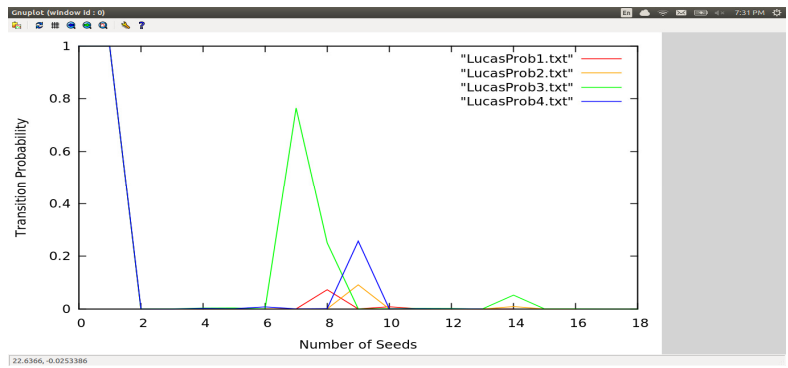

**Figure 4: The transition probability on VMs ($\frac{L_i}{L_n}$ versus n) for n=4, 8, 10, 12, j=8 and effects of tail-boosting**

**j = Number of dummy VM's for control of malicious seeds.**

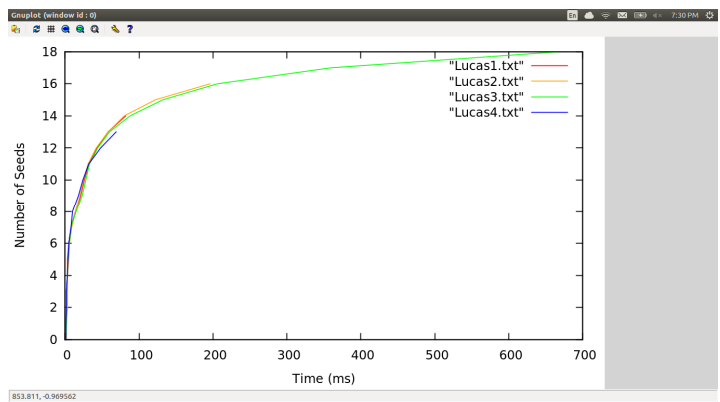

. **Figure 5:** $L_n$ **versus. Time--growth of seeds in time (milliseconds) for n=4, 8, 10, 12**

### 5. Conclusion

It is observed that, the seeds grow quickly and recursively, thereby warranting very little supervision. The numerical growth of the seeds is exponential, making it suitable to propagate in large numbers and rapid manner. The downside of the modeling technique is; if n increases, $L_n$ grows exponentially and consequently $\frac{L_i}{L_n}$ slows down. The probability that a new node might get attacked becomes less. The "Tail-boosting" dynamics achieves sharp decline of attack probability patterns but slows down the intensity of attacks gradually by scaling the network up. This is done by adding a bunch of "dummy VM's" which were injected into the network after the attack was initiated. We observe that, the probabilities eventually tend to zero implying the attack was nullified.

Different modeling techniques need to be discussed to help understand/combat the threat from attacking the systems including those of cloud. In this paper, we discussed the randomized recursive model and its consequent attack mechanisms by providing different equations using random Fibonacci sequences as well as Lucas equations. It is also indicated that malicious seeds grow exponentially with respect of these equations. The drawbacks of this approach also were taken into account. being recursive in nature provides defense mechanism to the attacks. The virtual machines (VMs) can be affected due to the fact that there is sharing of resource among these VMs. The magnitude of attack increases gradually until the entire system is affected. The computation of the recovery time is one metric the authors wish to find out and establish the relationship between the transition times and recovery times under such attack.

**Appendix A: Simulation Output Screenshots**

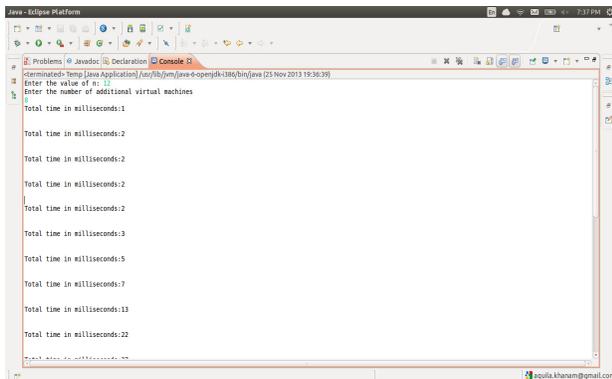

```
Total time in milliseconds:30

Total time in milliseconds:35

Total time in milliseconds:46

Total time in milliseconds:62

Total time in milliseconds:91

Total time in milliseconds:134

Total time in milliseconds:209

Total time in milliseconds:362

Total time in milliseconds:677

Total time in milliseconds:1315

Total time in milliseconds:2611
```

```
Total time in milliseconds:362

Total time in milliseconds:677

Total time in milliseconds:1315

Total time in milliseconds:2611

1.0
1.0
6.446398514416795E-11
6.306585098022473E-4
1.7393419115973085E-5
0.0040030029387403045
0.0011031113664338459
7.518300569222064E-5
5.916861468273144E-5
0.005245238275421209
1.609444130333043E-4
1.3711099401298292E-4
0.03378583950432645
2.0710733855471217E-6
1.4384067914357951E-4
0.0892995881460248
4.06998281852612113E-7
2.4478144389322695E-8
2.0803953034092535E-5
0.0
```